\documentclass[twocolumn,showpacs,preprintnumbers,amsmath,amssymb]{revtex4}

\usepackage{graphicx}% Include figure files
\usepackage{dcolumn}% Align table columns on decimal point
\usepackage{bm}% bold math

\begin{document}
\def\bsk{{\hbox{\boldmath $k$}}}
\def\vv{{\hbox{\boldmath $v$}}}

\title{Signals in Single-Event Pion Interferometry for
Granular Sources of Quark-Gluon Plasma Droplets}

\author{Wei-Ning Zhang$^{1}$}
\author{Shu-Xia Li$^1$}
\author{Cheuk-Yin Wong$^{2,3}$}
\author{M. J. Efaaf$^{1}$}

\affiliation{
$^1$Department of Physics, Harbin Institute of Technology, 
Harbin, 150006, P. R. China\\
$^2$Physics Division, Oak Ridge National Laboratory, Oak Ridge, TN
37831, U.S.A.\\
$^3$Department of Physics, University of Tennessee, Knoxville, TN
37996, U.S.A.
}

\date{\today}

\begin{abstract}
We investigate two-pion Bose-Einstein correlations of quark-gluon
plasma droplet sources in single-event measurements.  We find that the
distribution of the fluctuation between correlation functions of the
single- and mixed-events provide useful signals to detect the granular
structure of the source.
\end{abstract}


\maketitle

\section{Introduction}

Recently, there has been much progress in the experimental search for
the quark-gluon plasma \cite{Gyu04,Rik04}.  While there are many
indications suggesting the presence of a very dense matter produced in
high-energy heavy-ion collisions \cite{Gyu04,Rik04}, a very important
question is whether the produced dense matter is a quark-gluon plasma.
If it is a quark-gluon plasma, it will undergo a phase transition from
the quark-gluon plasma phase to the hadronic phase.  It is desirable
to search for the signature for the phase transition of the
quark-gluon plasma.

The signature for the phase transition however depends sensitively on
the order of the transition.  Previously, it was suggested by Witten
and many other workers that a granular structure of droplets occurs in
a first-order QCD phase transition, and the observation of the
granular structure can be used as a signature for a first-order QCD
phase transition \cite{Wit84,Sei89,Kaj91,Pra92,Cse92,Ven94,Zha95,Ala99,
Zha00,Cse03,Ran04,Zha04,Ran04a}.

In a recent analysis, a granular emitting source of droplets was put
forth to explain the HBT interferometry puzzle for nucleus-nucleus
collisions at RHIC \cite{Zha04}.  The suggestion was based on the
observation that in the hydrodynamical model \cite{Ris96,Ris98}, the
particle emission time scales with the radius of the droplet.
Particles will be emitted earlier if the radius of the droplet is
smaller, as in a source of many droplets.  An earlier emission time
will lead to a smaller extracted HBT radius $R_{\rm out}$. As a
result, the value of $R_{\rm out}$ can be close to $R_{\rm side}$ for
a granular quark-gluon plasma source \cite{Zha04}.  

Previously, Pratt $et~al.$ studied the HBT interferometry of granular
droplets by averaging over many events \cite{Pra92}.  Methods to
detect a granular structure by the single-event intensity
interferometry were recently proposed \cite{Won04}. It was found that
the single-event correlation function from a chaotic source of
granular droplets exhibits large fluctuations, with maxima and minima
at relative momenta which depend on the relative coordinates of the
droplet centers.  The presence of this type of maxima and minima of a
single-event correlation function at many relative momenta is a
signature for a granular structure and a first-order QCD phase
transition \cite{Won04}.

The difficulty of using the single-event two-pion interferometry at
RHIC arises because of the small number of pion pairs with small
relative momenta.  In a typical single event of a nearly head-on
collision at very high energies at RHIC, the number of identical pions
is of the order of a few thousand.  The number of observed identical
pions $n_\pi$ is only a small fraction of this number.  For example,
the number of identical pions detected in the STAR Collaboration in
the most central Au-Au collisions at RHIC is of the order of a few
hundred \cite{Ada03}.  Although the number of pairs of identical pions
in the event varies as $N_{\pi \pi}=n_\pi(n_\pi-1)/2$, only a small
fraction of these pairs have relative momentum small enough to be
useful in a HBT analysis.  The number of pion pairs in each relative
momentum bin may be so small that there can be large associated
statistical errors.

Instead of trying to obtain the detailed granular structure of the
emitting source in each event at present, it will be useful in the
initial stage to have a more modest goal.  It is desirable to see
whether the correlation functions indicate possible signals for a
granular structure. 
More refined study of the granular structure can follow after the
initial stage becomes successful.

Accordingly, we shall try to outline a method to extract the signals
for the granular structure.  Our idea is to calculate the fluctuations
of the single-event correlation function relative to its corresponding
mixed-event correlation function.  The difference constitutes the
``signal'' for the event in question.  The distribution of these
fluctuations (in units of their statistical errors), collected for a
large number of single-events to enhance statistics, would have a
wider distribution for a granular structure, compared to those from
emitting source without the granular structure.  The distribution of
the correlation function fluctuations provide a useful tool to detect
the granular structure of the source.  In particular, the
root-mean-square fluctuation, in the case of a granular droplet
structure, increases when the number of droplets decreases.  If the
phase transition is accompanied by only a few number of droplets, the
signal may be large enough to make it detectable.  In what follows, we
would like to analyze whether this method may be feasible.

\section{Single-event and mixed-event two-pion correlation functions}

The two-particle Bose-Einstein correlation function for the detection
of identical pions with laboratory momenta $k_1$ and $k_2$ is defined
as $C(k_1,k_2)=P(k_1,k_2)/P(k_1)P(k_2)$ where $P(k_1,k_2)$ is the
two-particle momentum distribution and $P(k_i)$ is the single-particle
momentum distribution with momentum $k_i$.  For a chaotic
pion-emitting source, $P(k_i)~(i=1,2)$ is
\begin{eqnarray}
\label{pk1}
P(k_i) = \sum_{X_i} A^2(k_i,X_i) \,,
\end{eqnarray}
where $A(k_i,X_i)$ is the magnitude of the amplitude for emitting a
pion with 4-momentum $k_i=(E_i,\bsk_i)$ at $X_i$.  The two-particle
distribution function $P(k_1,k_2)$ can be expressed as
\begin{eqnarray}
\label{pk12}
P(k_1,k_2) = \sum_{X_1, X_2} \Big|\Phi(k_1, k_2; X_1,
X_2 )\Big|^2 ,
\end{eqnarray}
where $\Phi(k_1, k_2; X_1, X_2 )$ is the two-pion wave function.
Neglecting the absorption of the emitted pions by other droplets,
$\Phi(k_1, k_2; X_1, X_2 )$ is simply \cite{Won94}
\begin{eqnarray}
\label{PHI}
& &\!\!\!\!\!\!\!\Phi(k_1, k_2; X_1, X_2 )
\nonumber \\
& & ={ \frac{1}{\sqrt{2}} } \Big[ A(k_1, X_1)
A(k_2, X_2) e^{i k_1 \cdot X_1 + i k_2 \cdot X_2}
\nonumber \\
& & ~~~+ A(k_1, X_2) A(k_2, X_1)
e^{i k_1 \cdot X_2 + i k_2 \cdot X_1 } \Big] .~~~~~
\end{eqnarray}

The correlation function $C(k_1,k_2)$ is in general a function of the
4-dimensional momentum $k_1-k_2$ and $k_1+k_2$.  Previous results of
the single-event correlation function of granular droplets indicate
large fluctuations as a function of the relative 4-momentum
$k_1-k_2$, having maxima and minima at locations which depend on the
relative coordinates of the droplet centers \cite{Won04}.  To map out
the details of such a multi-dimensional correlation function, it is
necessary to have a large number of pion pairs in a single event which
may be beyond the capabilities of present detectors and accelerators.
With limited statistics as would likely be the case, we can first
study the simplifying case by concentrating on a small number of
degree of freedom and integrating out other degrees of freedom so that
the statistical errors in the correlation function can be smaller.
For this purpose, we shall study the correlation function as a
function of the variable $q=|{\bf k_1}-{\bf k_2}|$ for which the other
degrees of freedom have been integrated out.

In our numerical work for a granular source with $N_d$ droplets, we
obtain the single-event and mixed-event two-pion correlation functions
for the granular source with the following steps:

Step 1: Generate the space-time coordinates $R_j~(j=1,2,\dots,N_d)$ of 
the droplet centers according to a distribution of the droplet centers. 

Step 2: Select the two emitting-pions from the droplets randomly, and
get the space-time coordinates $X_1$ and $X_2$ of the pions in the
laboratory frame according to the density distribution of the
droplets, the coordinates of the droplet centers, and the collective
velocities at the emission points $\vv(X_i)~(i=1,2)$.

Step 3: Generate the momenta $k'_1=(E_1',{\bf k}_1')$ and
$k'_2=(E_2',{\bf k}_2')$ of the two pions in the frame in which the
source element is at rest according to the distribution $A^2(k_i',X)$,
taken to be the Bose-Einstein distribution characterized by the
temperature $T_f$, and obtain their momenta $k_1$ and $k_2$ in the
laboratory frame by Lorentz transformation if the source element is
boosted in the laboratory frame.
                                                                               
Step 4: Accumulate the event in the bin of the corresponding relative
momentum variable $q=|{\bf k_1}-{\bf k_2}|$ with the weight factor 
$w_{11}$ for the probability $P(k_1)P(k_2)$ for a pair of uncorrelated 
pions of relative momentum $q$, and with the weight factor $w_{12}$ for 
the probability $P(k_1,k_2)$ for a pair of correlated pions of relative 
momentum $q$, 
\begin{eqnarray}
\label{w11}
w_{11} = [E'_1/E_1][E'_2/E_2] \,,
\end{eqnarray}
\begin{eqnarray}
\label{w12}
w_{12} &=&{ \frac{1}{2}} \Bigg|\sqrt{E'_1/E_1}\sqrt{E'_2/E_2}~
e^{i k_1 \cdot X_1 + i k_2 \cdot X_2}
\nonumber \\
&+ &\sqrt{E'_{12}/E_1}\sqrt{E'_{21}/E_2}~
e^{i k_1 \cdot X_2 + i k_2 \cdot X_1 } \Bigg|^2 \,, 
\end{eqnarray}
where $E'_{ij}~(i,j=1,2)$ is the energy of the $i-$th pion in the
frame in which the source element at $X_j$ is at rest, which is
obtained from $E_i$ by a reverse Lorentz transformation with the
collective velocity $\vv(X_j)$.

Step 5: Repeat steps 2 through 4 for $N_{\pi\pi}$ pairs of pions in a
single event.  We label the distributions obtained for $P(k_1)P(k_2)$
for a pair of uncorrelated pions by ${\rm Uncor}_s (q)$ and the
distribution of $P(k_1,k_2)$ for a pair of correlated pions by ${\rm
Cor}_s (q)$.

Step 6: Repeat steps 1 through 5 for $N_{\rm event}$ number of
different events, and obtain the mixed-event correlated and
uncorrelated pion-pair distributions ${\rm Cor}_m (q)$ and ${\rm
Uncor}_m (q)$, by summing ${\rm Cor}_s (q)$ and ${\rm
Uncor}_s (q)$ of the $N_{\rm event}$ different events.

Step 7: Obtain the single-event correlation function $C_s(q)$ by dividing 
${\rm Cor}_s (q)$ by ${\rm Uncor}_m (q)$ and the mixed-event correlation 
function $C_m(q)$ by dividing ${\rm Cor}_m (q)$ by ${\rm Uncor}_m (q)$. 

We first investigate the two-pion correlation functions for $N_d$
static granular droplet sources.  The centers of the droplets are
assumed to follow a Gaussian distribution with a standard deviation
$\sigma_R$, and the density distribution of each droplet is assumed to
be given by a Gaussian distribution with a standard deviation
$\sigma_d$.  In our numerical examples, $\sigma_R$ and $\sigma_d$ are
taken as 5.0 and 1.5 fm, and the thermal emission temperature of the
pions is taken to be $0.65 T_c= 0.65\times160=104$ MeV.  Fig. 1 (a),
(b), and (c) show the two-pion correlation functions for the granular
sources with $N_d=$4, 8, and 16, respectively.  In each figure the
dashed lines give the correlation function $C(q)$ for a sample of
different single events each of which has a correlated pion-pair
distribution ${\rm Cor}_s (q)$ calculated with $N_{\pi\pi}=10^6$ pion
pairs within $q\leq250$ MeV, and the solid line is for the mixed-event
obtained by averaging $N_{\rm event}=10^3$ single events.  One can see
that there are fluctuations for the single-event correlation functions
relative to the mixed-event correlation function, and the fluctuations
increase as $N_d$ decreases.

\begin{figure}
\includegraphics[angle=0,scale=0.9]{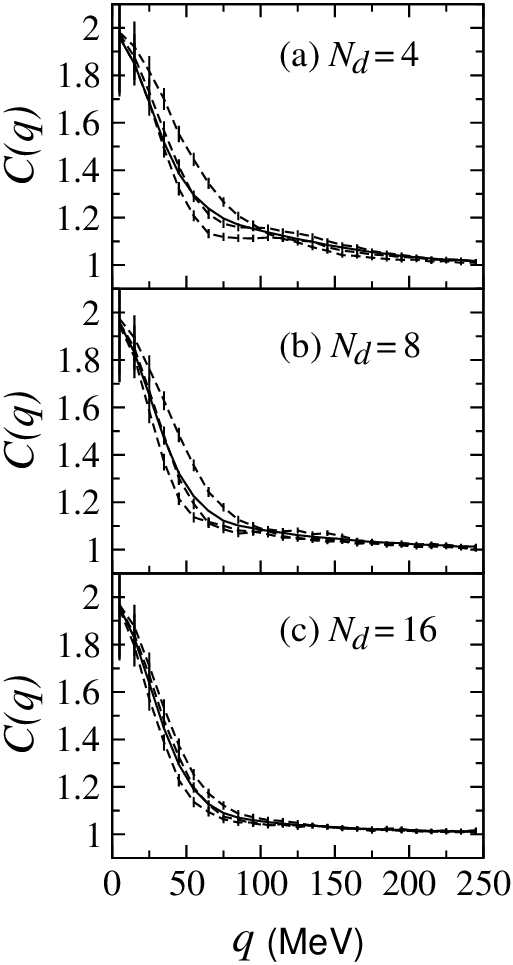}
\caption{\label{fig:f1} Two-pion correlation functions for a sample of
different single events (dashed lines) and mixed events (solid
lines) for static granular sources of $N_d$ droplets.}
\end{figure}

\section{Distribution of the fluctuation of single-event two-pion 
correlation function}

Form the results in Fig.\ 1, we observe that the correlation function
$C_s(q)$ for individual single events fluctuates with respect to the
mixed-event correlation function $C_m(q)$.  We can make the
fluctuation quantitative and introduce the fluctuation as the
difference between the single-event correlation function and the
mixed-event correlation function.  In order to take into account the
error of the measurement, we weigh the fluctuation by the inverse of
the corresponding error and define the fluctuation quantitatively as
\begin{eqnarray}
\label{RF}
f(q)  = \frac{| C_s(q) - C_m(q) |}{\Delta | C_s(q) - C_m(q)|} \,.
\end{eqnarray}
where $\Delta | C_s(q) - C_m(q)|$ is the error in the measurement of 
$C_s(q)-C_m(q)$ given by
\begin{eqnarray}
\label{Error}
& &\hspace*{-5mm} \Delta |C_s(q)- C_m(q)|=\Delta C_s(q)+\Delta C_m(q) 
\simeq \Delta C_s(q) 
\nonumber \\ 
& &
\simeq C_s(q) \Bigg\{ { \frac{1}{ {\sqrt{N_{\pi\pi}} }}} 
+ { \frac{1} {{\sqrt{{\rm Cor}_s(q)} }}} \Bigg\} . \hspace*{4mm}
\end{eqnarray}

\begin{figure}
\includegraphics[angle=0,scale=0.6]{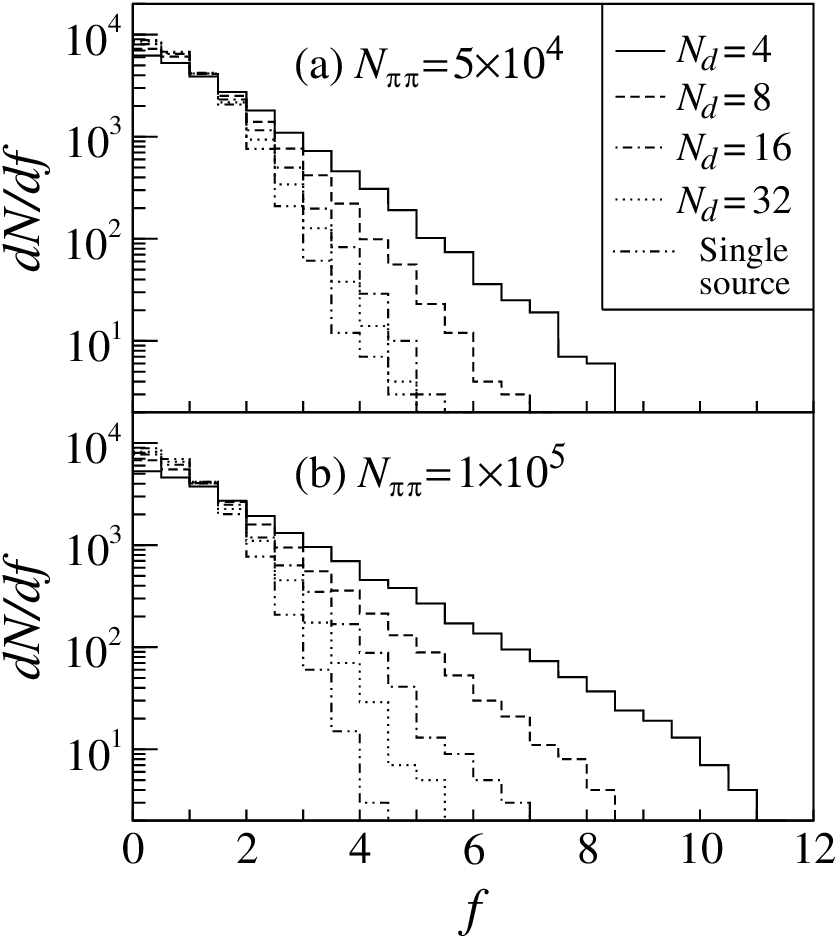}
\caption{\label{fig:f2} The distributions $dN/df$ of $f$ for (a)
$N_{\pi\pi}=5\times10^4$ and (b) $N_{\pi\pi}=10^5$.  
}
\end{figure}

We calculate the distribution of the fluctuation $f$ for the sources
with different numbers of droplets. In these calculations, we take the
width of the relative momentum bin as 10 MeV and use the bins in the
region $20\leq q\leq 250$ MeV.  Fig. 2 (a) and (b) show the
distributions of $f$ for $N_d=4, 8, 16,{\rm~ and~} 32$, obtained from
1000 single events each of which has the correlated pion-pair
distribution ${\rm Cor}_s (q)$ calculated with
$N_{\pi\pi}=5\times10^4$ and $10^5$ pion pairs within $q\leq250$
MeV.  The standard deviations $\sigma_R$ and $\sigma_d$ for the
granular sources are 5.0 and 1.5 fm.  The results for a non-granular
single source with a Gaussian density distribution with 5.0 fm standard
deviation are also shown as a reference.  It can be seen that the
distributions for the granular sources are wider than that for the
single source.  The width of the distribution for the granular
source decreases with $N_d$.

From the distribution of $f$, one can calculate the root-mean-square
values of $f$.  Fig. 3 (a) and (b) show the root-mean-square $f$ as a
function of $N_d$ for sources with $\sigma_R=$5.0 and 7.0 fm obtained
for for the cases of $N_{\pi\pi}=10^5$ and $\sigma_d=$1.5 fm, and
$N_{\pi\pi}=5\times10^4$, and $\sigma_d=$2.0, 1.5, and 1.0 fm.  The
double-dot-dashed lines are the root-mean-square $f$ for the
distribution of the single Gaussian source in Fig. 2.  It can be seen
that $f_{\rm rms}$ is sensitive to $N_d$ and decreases as $\sigma_d$
and $\sigma_R$ increase.  With the values of $f_{\rm rms}$, one may
distinguish the granular source with $N_d$=16 from the single Gaussian
source with $N_{\pi\pi}=5\times10^4$ pairs of identical pions, and one
may even distinguish the granular sources up to $N_d=32$ with
$N_{\pi\pi}=10^5$ identical pions.

\begin{figure}
\includegraphics[angle=0,scale=0.7]{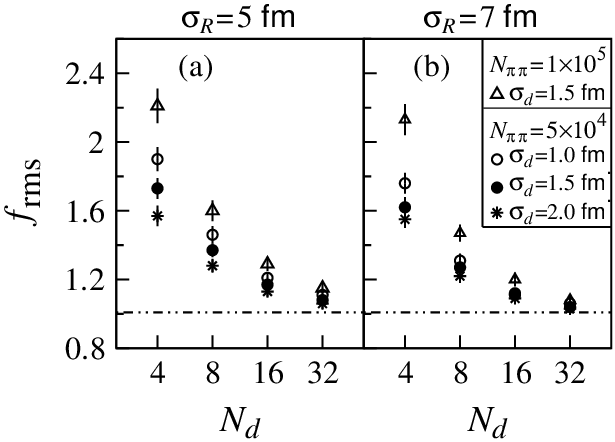}
\caption{\label{fig:f3} The root-mean-square $f$ as a function of
$N_d$ for static granular sources.  The double-dot-dashed lines are
the result for a single Gaussian source.  }
\end{figure}

\section{Distribution of the fluctuation for hydrodynamical QGP 
droplet source}

\begin{figure}
\includegraphics[angle=0,scale=0.9]{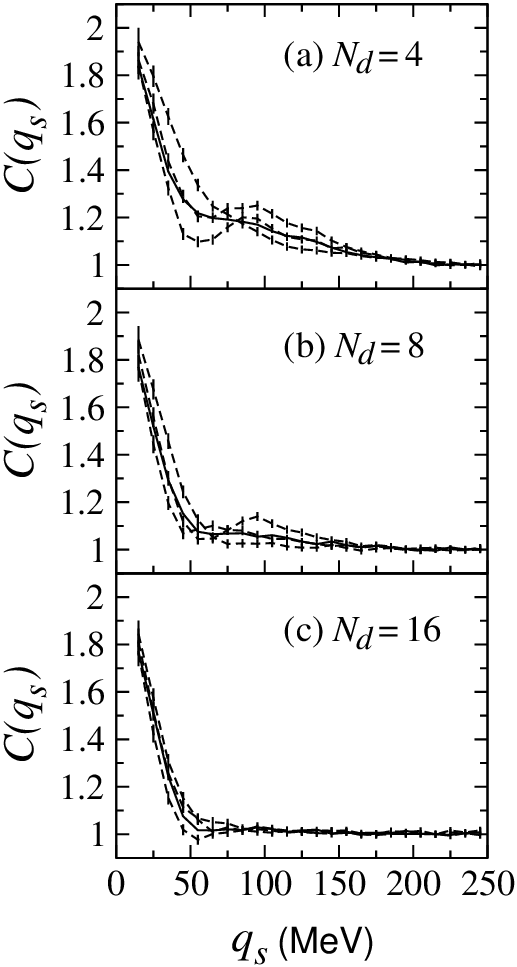}
\caption{\label{fig:f4} Two-pion correlation functions for a sample of 
different single-events (dashed lines) and mixed events (solid
lines) of the dynamical granular sources.}
\end{figure}

We investigate next the distribution of $f$ for a granular sources of
quark-gluon plasma droplets which evolve hydrodynamically.  We assume
that all of the droplets in the source have the same initial radius
$r_d=1.5$ fm and evolve hydrodynamically in the same way.  We use
relativistic hydrodynamics with the equation of state of the entropy
density \cite{Bla87,Lae96} to describe the evolution of the droplets 
\cite{Ris96,Ris98,Zha04}, and take the temperature width of the transition 
as $\Delta T=0.05$ and the initial conditions of the droplets as 
\cite{Ris96,Ris98,Zha04}
\begin{eqnarray}
\epsilon'(t'=0,r') = \left\{
           \begin{array}{ll}
           \epsilon_0, & r'<r_d, \nonumber\\
           0, & r'>r_d,
           \end{array}  \right. 
\nonumber
\end{eqnarray}
\begin{eqnarray}
\label{inic}
v'(t'=0,r') = \left\{
           \begin{array}{ll}
           0, & r'<r_d, \\
           1, & r'>r_d,
           \end{array}  \right.
\end{eqnarray}
where, $r'$, $\epsilon'$, and $v'$ are the radial coordinate, energy
density, and velocity of a fluid element in the droplet-rest frame,
$\epsilon_0=1.875 T_c s_c$ \cite{Ris96,Ris98} is the initial energy
density of the droplets.  The initial distribution of the droplet
centers is taken to be a Gaussian distribution with the standard
deviation $\sigma_R=5.0$ fm.  For the case with an additional
collective radial expansion, the droplet centers are assumed to have a
constant radial velocity $v_d$ in the center-of-mass frame of the
granular source \cite{Zha04}.  In order to reduce the influence of
source lifetime on our observations for the granular structure
\cite{Pra92}, we use the ``side'' component of the relative momentum
of the two pions $q_s$ (perpendicular to the total momentum of the two
pions) as the variable \cite{Pra90,Ber88,Wie99,Wei02} for the granular
source of hydrodynamic evolution QGP droplets.  Fig. 4 (a), (b), and
(c) show the correlation function $C(q_s)$ (with $q_{\rm out} \leq 20$
MeV) for the dynamical granular sources with $v_d=0.5$ and $N_d=$4, 8,
and 16, respectively.  In each figure the dashed lines give the
correlation function $C(q_s)$ for a sample of different single events
each of which has a correlated pion-pair distribution ${\rm Cor}_s
(q_s)$ calculated with $N_{\pi\pi}=10^6$ pion pairs within $q_s
\leq250$ MeV, and the solid line is for the mixed-event obtained by
averaging $N_{\rm event}=10^3$ single events.  The freeze-out
temperature of the pions is taken to be $0.65T_c = 0.65\times160=104$
MeV.

Fig. 5 (a) and (b) show the distributions of $f$ for the dynamic
granular sources with $v_d=0$ and $v_d=0.5$, and $N_d=4$, $N_d=8$,
$N_d=16$, and $N_d =32$.  The double-dot-dashed lines are for a
dynamical single source with initial radius $r_d=$5.0 fm as a
reference.  The number of pion pairs within $q_s\leq250$ MeV for one
single event is $N_{\pi\pi}=5\times10^4$ and the number of events is
$N_{\rm event}=10^3$ for all the distributions.  Because of the radial
expansions of the droplets, the two pions emitted from the
single-droplet source are boosted along the radial direction of the
source and the two pions emitted from the granular source are boosted
along random direction when $v_d=0.5$.  It leads to a smaller width of
the distribution for the single-droplet source than those for the
granular sources.  When $v_d \ne 0$ the widths of the distributions
for the granular sources decrease because of the additional collective
radial expansion $v_d$.

\begin{figure}
\includegraphics[angle=0,scale=0.5]{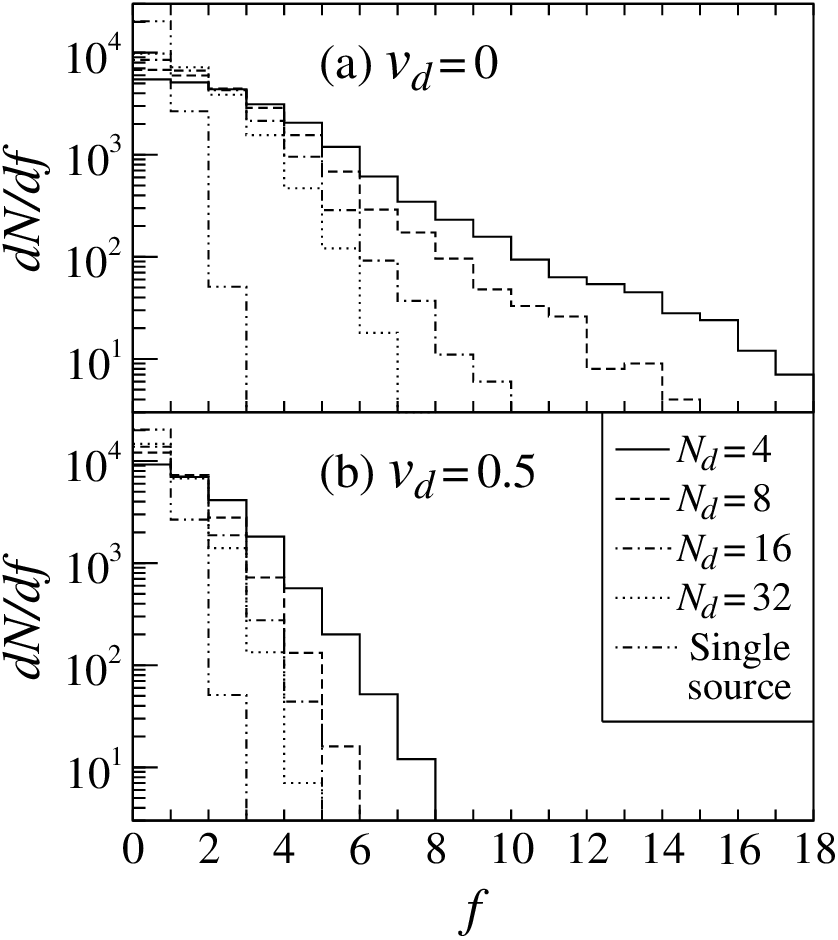}
\caption{\label{fig:f5} The distributions of $f$ for the dynamical
quark-gluon plasma granular sources with $v_d=$0 and 0.5.}
\end{figure}

\begin{figure}
\includegraphics[angle=0,scale=0.7]{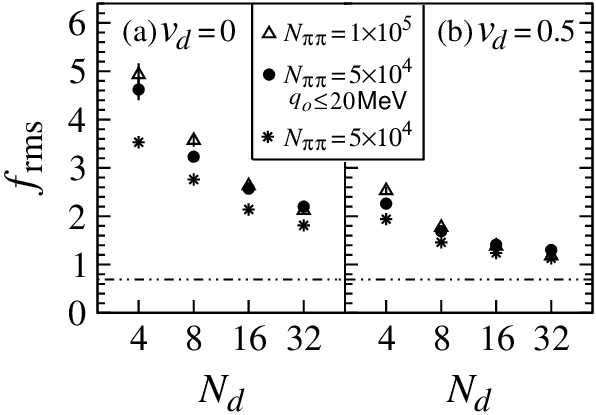}
\caption{\label{fig:f6} The root-mean-square $f$ as a function of
$N_d$ for dynamical granular sources.}
\end{figure}

Fig. 6 (a) and (b) show the root-mean-square values of $f$ calculated
from the distributions of one thousand single-events for the dynamical
granular sources with $v_d=$0 and 0.5, respectively.  The symbols of
{\scriptsize $\triangle$} are for the case of the number $N_{\pi\pi}$
of the pion-pair within $q_s\leq250$ MeV for one single event is
$10^5$.  The symbols of $\bullet$ and $\ast$ are for the cases of the
$N_{\pi\pi}=5\times10^4$ and with and without the constraint $q_o \leq
20$ MeV.  The double-dot-dashed lines are the result for the
distribution of the single-droplet source in Fig. 5.  One observes
that $f_{\rm rms}$ for the granular sources are larger than that for
the single-droplet source, and the root-mean-square $f$ for the
granular sources increase with $N_{\pi\pi}$.  Although the $f_{\rm
rms}$ results with the constraint $q_o \leq 20$ MeV are larger than
those without the constraint, our simulations indicate that this
constraint is very restrictive and only about 3.5 percent of the pion
pairs within $q_s\leq250$ MeV satisfy it.  As the droplet number
decreases, the fluctuation increases.  The effect of the increase
becomes less pronounced as the collective expansion velocity
increases.

\section{Conclusions and Discussions}

Recent experiments at RHIC provide ample evidence for a dense matter
produced in high-energy heavy ion collisions.  Is the produced dense
matter the quark-gluon plasma?  If so, what is the order of its phase
transition?  As a granular structure of droplets occurs in a
first-order QCD phase transition, the observation of the granular
structure can be used as a signature for a first-order QCD phase 
transition \cite{Wit84,Sei89,Kaj91,Pra92,Cse92,Ven94,Zha95,Ala99,
Zha00,Cse03,Ran04,Zha04,Ran04a}.

We would like to develop tools to use HBT interferometry to examine
the granular droplet structure of the dense matter if they are
produced in high-energy heavy-ion collisions.  We show previously that
the single-event correlation function in HBT interferometry for
granular droplets exhibits oscillations, depending on the relative
coordinates of the droplet centers \cite{Won04}.  In realistic
experimental situations, the number of identical pion pairs in each
single event is limited. We continue our investigation here to find
out appropriate measurable quantities that could be used to detect
granular structure in HBT measurements.

In our present investigation, we study the fluctuation between the
single-event correlation function and the mixed-event correlation
function and find that the distribution of the correlation function
fluctuation $f$ between the single-event and the mixed-event
correlation functions can be a measurable quantities that could be
used to probe the granular droplet structure.  The width of the
distribution is greater for a granular source than that for a single
source, and the width increases as the droplet number decreases.  The
effect of the increase becomes less pronounced when the droplets have
a collective expansion.  These changes of the widths can be quantified
in terms of the root-mean-square fluctuation of $f$.  The $f_{\rm
rms}$ for a granular droplet source increases with the number of
identical pion pairs in an event.  The detection of the granular
droplets becomes more favorable as the number of identical pairs
increases.

Our numerical simulated calculations indicate that the correlation
function fluctuation leads to detectable differences if the droplet
number is small (less than or equal to 16) and the number of identical
pairs in each single event is of order $5\times 10^4$ or more.  At
RHIC energies, the multiplicity of identical pion of event is about a
few hundreds, and the number of pion-pairs in an event is about
$10^5$.  It will be of interest to see whether the correlation
function fluctuation can indeed be measured at RHIC.  The situation
become even more favorable for LHC collisions at higher energies where
there can be a greater number of identical pion pairs.

As with the development of many new experimental tools, progress is
made by gradually increasing the complexity of one's scopes of
research and the areas of focus.  In the present analysis, we have not
considered the absorption and the multiple scattering of the pions
\cite{Won02,Zha04a}.  Investigations on these effects of can be
carried out in the future to see how they may modify the distribution
of the fluctuations.  We have also not considered the fluctuation of
the overall size of emission source which clearly depends on the
experimental selection and will require an investigation in
conjunction with the experimental set-up and selections.  An event
multiplicity cut may be needed to reduce the fluctuation due to the
sizes of the emitting source.  Future investigations to refine the
tool of HBT measurements for the detection of granular structure will
be of great interest to probe the order of the quark-gluon plasma
phase transition.

\begin{acknowledgments}
This research was supported
by the National Natural Science Foundation of China under Contract
No.10275015 and by the Division of Nuclear Physics, US DOE, under
Contract No. DE-AC05-00OR22725 managed by UT-Battle, LC.
\end{acknowledgments}

\end{document}